\documentclass[runningheads]{llncs}
\usepackage[T1]{fontenc}
\usepackage{graphicx}
\usepackage{listings}

\begin{document}
\title{Workflows to driving high-performance interactive supercomputing for urgent decision making}
\titlerunning{Workflows for urgent high-performance interactive supercomputing}
%
\author{Nick Brown\inst{1} \and Rupert Nash\inst{1} \and Gordon Gibb\inst{1} \and Evgenij Belikov\inst{1} \and Artur Podobas\inst{2} \and Wei Der Chien \inst{2} \and Stefano Markidis \inst{2} \and Markus Flatken\inst{3} \and Andreas Gerndt\inst{3}}
\authorrunning{N. Brown et al.}
%
\institute{EPCC, The University of Edinburgh, Edinburgh, UK \and
KTH Royal Institute of Technology, Stockholm, Sweden\and
German Aerospace Center (DLR), Braunschweig, Germany}
\maketitle              
\begin{abstract}
Interactive urgent computing is a small but growing user of supercomputing resources. However there are numerous technical challenges that must be overcome to make supercomputers fully suited to the wide range of urgent workloads which could benefit from the computational power delivered by such instruments. An important question is how to connect the different components of an urgent workload; namely the users, the simulation codes, and external data sources, together in a structured and accessible manner. 

In this paper we explore the role of workflows from both the perspective of marshalling and control of urgent workloads, and at the individual HPC machine level. Ultimately requiring two workflow systems, by using a space weather prediction urgent use-cases, we explore the benefit that these two workflow systems provide especially when one exploits the flexibility enabled by them interoperating.
\end{abstract}
\section{Introduction}

From human health emergencies to natural disasters, the global pandemic and the recent bouts of extreme climate events have demonstrated the need to make urgent, accurate, decisions for complex problems. The use of near real time detection of unfolding disasters and computational modelling of such situations is a powerful tool in aiding urgent responders to tackle such disasters and disease outbreaks. Combining HPC computational models with real-time data and interactive user interaction can significantly aid in such urgent decision-making for disaster response and other societal issues, which ultimately saves lives and reduces economic loss. 

However the major challenge is that whilst HPC machines have a long tradition of simulating disasters in retrospect, they have not been commonly used \emph{in-the-loop} whilst a disaster is unfolding in real-time. There are numerous reasons for this including limits imposed based upon the classical way in which users interact with HPC machines via the batch queue system. Recent years have seen numerous advances in technologies and machine access policies that open up the possibility of using such HPC machines in a more interactive fashion for urgent workloads, and a major question is how we should best develop our codes to most effectively exploit these technologies.

In this paper we explore the role of workflows in high-performance interactive supercomputing for supporting urgent decision making. The paper is structured as follows, after briefly describing the background to this work in Section \ref{sec:background} we then explore the use of our workflows in Section \ref{sec:workflows}. Section \ref{sec:casestudy} uses a space weather prediction urgent workload to explore key facets of our approach on ARCHER2, the UK national supercomputer, before drawing conclusions and discussing further work in Section \ref{sec:conclusions}.

\section{Background}
\label{sec:background}

There are numerous examples of emergency situations that we face as society including COVID, wildfires, hurricanes, extreme flooding, earthquakes, tsunamis, winter weather conditions, public unrest, food and energy resource management, and traffic accidents. Numerical modelling has already demonstrated \cite{mosquito} \cite{wildfire} \cite{cheese} that it can contribute to insights which will then benefit these areas. However this alone is not enough, as in order to drive these codes for such workloads then the ability to consume real-time input data \cite{role_interactive}, and enable interaction with emergency responders in the field using rich visualisation technologies such as ParaView \cite{paraview} is required. 

However a key challenge is that there are numerous facets at play, often comprising multiple simulation codes that might need to be coupled in non-trivial ways, numerous data sources which might publish new data at unpredictable times when it becomes available (and which might then require the creation of new simulation instances), and the end-users who need to interact with running simulations or view processed results. An important question is how to connect all of these together in the most efficient manner which is complicated by the fact that we aim to support a general solution that can be applied to many different urgent workloads. We have found that the answer to this is the use of workflows however there is not a simple \emph{one size fits all approach}, but instead different workflow technologies suit different parts of the technology stack.

Workflows are highly popular in other fields such as bioinformatics \cite{cwl_bio}, and are becoming steadily more popular in HPC especially as the community continues to embrace data workloads. There are numerous workflow technologies and choices, with the most popular including the Common Workflow Language (CWL) \cite{cwl}. Nevertheless, there is still work to be done in successfully exploiting many of these in the field of supercomputing.

\section{The anatomy of our workflows}
\label{sec:workflows}
\begin{figure}[h]
  \centering
  \includegraphics[scale=0.67]{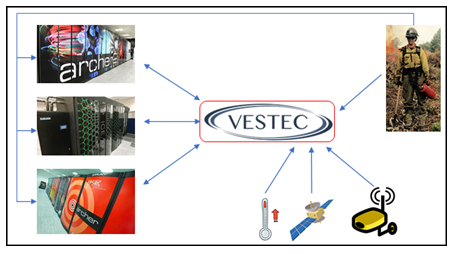}
  \caption{Overview of VESTEC system and interaction with users, data sources, and HPC machines.}
  \label{fig:vestec_architecture}
\end{figure}

In our approach we deploy workflows in two major areas, and these are illustrated in Figure \ref{fig:vestec_architecture}. The first is in our marshalling and control system which is a standalone system, represented by \emph{VESTEC} in Figure \ref{fig:vestec_architecture}, and drives the execution of workloads across the HPC machines. The second is on the HPC machines themselves where the simulation codes and supporting functionality will actually execute.

\subsection{Marshalling and control system workflows}
\label{sec:marshallcontrol}
The VESTEC marshalling and control system drives the execution of workloads across HPC machines. It is not installed on a supercomputer, as the system is not intended to undertake any computationally intensive tasks. Instead, this middleware technology resides on a server, most likely enterprise class, and will manage the life span of urgent workloads when responding to specific disasters. 

\begin{figure}[h]
  \centering
  \includegraphics[scale=0.57]{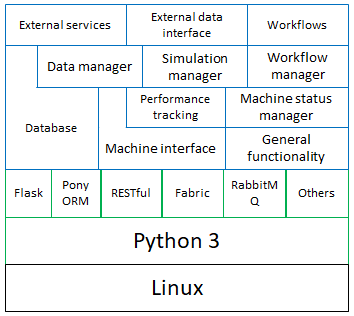}
  \caption{Illustration of technology stack of the marshalling and control system}
  \label{fig:stack}
\end{figure}

Figure \ref{fig:stack} illustrates the technology stack view of our marshalling and control system, where the two black boxes on the bottom layers represent support required by the hosting server that the system is running on, namely Linux OS and Python. In green are a subset of the major Python packages that are in use by the system. The blue boxes above represent constituent components of the VESTEC system, where at the top the \emph{external services} presents a publicly accessible API for clients to integrate with the system for the management of incidents, users, and the system itself. The \emph{external data interface} enables the system to both poll for new data from sources such as sensors, and for external sources, such as client GUIs, to push data into the system. 

However the major way in which urgent workload owners integrate with the system is by developing their own workflows definitions and plugging these into the system which is represented by \emph{workflows} in Figure \ref{fig:stack}. Our view is for lower layers of the system stack to provide a series of services and managers that undertake specific activities required by the workflows and can be called using a well documented API. Put simply, we provide a separation of concerns where lower layers of the stack provide the mechanism that urgent workload owners can leverage by developing their bespoke workflows which constitute the policy side of what such workloads require. 

Moreover, we have found that it is effective to present the marshalling and control system to users as primarily a workflow system, where each workflow stage represents progression through a disaster’s lifetime and these stages are triggered by some combination of external stimulus and/or preceding workflow stages. Consequently, to integrate a disaster scenario with the system then one must develop a workflow description. Individual stages can undertake a wide range of functionality including data transformation, preparation and submission of job to an HPC machine, and data clean-up activities. At any point during execution, stages can send messages to corresponding queues which will activate other stages. We provide in our \emph{workflow manager} the building blocks required for users to express workflows, and Listing \ref{lst:workflow} illustrates a sketch of workflow code where programmers are interacting with the \emph{simulation manager} to submit a job to an HPC machine. There are two phases required for job submission, firstly the creation of the job which determines which machine to allocate to \cite{mosquito} and creates the necessary folders, and secondly the submission of the job to the batch queue system. The reason for this multi-phase approach is that in the middle the workflow code can then create, copy, or move data to this location before submission, for example in Listing \ref{lst:workflow} the \emph{data manager} is called to put some configuration onto the HPC machine between lines 10 and 15.

\begin{lstlisting}[frame=lines,caption={Sketch of workflow code required for creating and submitting a simulation on an HPC machine},label={lst:workflow}, numbers=left]
try:
  callbacks = { 'COMPLETED': callback }
  sim_id = createSimulation(incidentID, 120, "00:15:00", "Example   
                 simulation", "submit.sh", callbacks, 
                 template_dir="templates/mysimulation")

  simulation=Simulation[sim_id]    
  machine_name=simulation.machine.machine_name            

  try:                
    putByteDataViaDM("myconfig", machine_name, "Simulation   
                     configuration", "text/plain", configuration_data, 
                     path=simulation.directory)                 
  except DataManagerException as err:
    print("Can not write simulation configuration "+err.message)        

  submitSimulation(sim_id)
except SimulationManagerException as err:
  print("Error creating or submitting simulation "+err.message)      
\end{lstlisting}

\begin{figure}[h]
  \centering
  \includegraphics[scale=0.57]{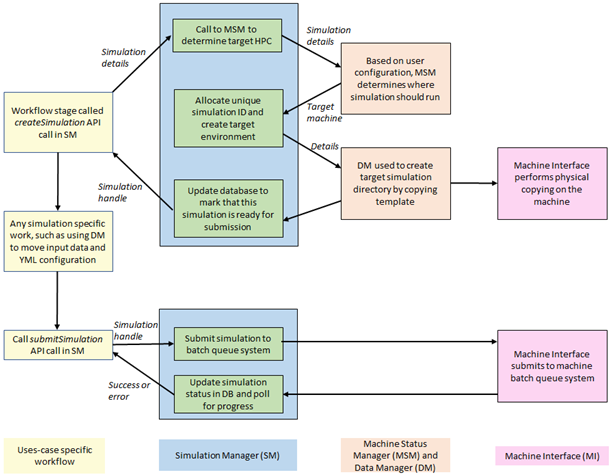}
  \caption{Illustration of how the workflow creates and submits jobs on the HPC machines by calling into the simulation manager which itself will call other appropriate internal services}
  \label{fig:workflow}
\end{figure}

The overarching flow and interaction with underlying services for this example is illustrated in Figure \ref{fig:workflow}, where it can be seen that the user's workflow code (on the left in yellow) calls into the \emph{simulation manager} which itself will then issue calls to other parts of the technology stack into order to undertake the required activities. This call is non-blocking, where once job submission has completed and the job is waiting in the queue then control flow will return to the workflow, and once the job reaches a successful completion stage then the callback workflow stage will be executed. This is provided via the \emph{callbacks} dictionary in Listing \ref{lst:workflow}, again leveraging the workflows concept, to execute a workflow stage provided by the user once a job has reached a specific state of execution (in this case completion, but it can also be when the job has started to run or an error has occurred).

The purpose of this paper is not to describe the underlying services or activities in detail, as there are many such processes involved in the system but instead to illustrate how our approach provides overarching infrastructure for the development and management of workflows via the \emph{workflow manager}, which then call into a set of internal services via their APIs in order to undertake specific actions on the HPC machines. 

\subsection{HPC machine side workflows}

In Section \ref{sec:marshallcontrol} we described the role of workflows from the VESTEC middleware perspective which provide marshalling and control functionality. Additionally, workflows are also useful on the HPC machines themselves to support interactive urgent workloads. The initial reason for this was to enable coupling of applications on an HPC machine, where the results from one application feed in as input to a subsequent code which is run when the former terminates. This requirement for coupling is commonplace, for instance undertaking pre-processing before execution of the main simulation code or post-processing of results. In this work we have adopted the Common Workflow Language (CWL) \cite{cwl} which is a specification for workflows common in fields such as bioinformatics. There is a CWL reference runner tool, which is used in this work, to drive these workflows and in previous work this was extended to increase compatibility with expressing MPI workloads \cite{cwl-mpi}. The intention of using CWL has been to describe the simulation steps on the HPC machine as workflow stages in CWL, with the reference runner tool then submitted to the batch queue system and executing the workflow on the allocated nodes. We found that the benefits of driving the HPC simulation jobs via CWL include:

\begin{itemize}
    \item Being able to inject configuration options that are specific to a disaster, scenario, or HPC machine via YAML configuration file(s). The overarching CWL workflow, that has been provided as a skeleton, is then concretised, or fleshed out, based upon this information.
    \item The ability to write a generic workflow description for a disaster only once which is independent across HPC machines. Whilst some machine configuration specifics might need to be provided, these can sometimes be shared across use-cases and/or represent a small number of machine specific tuning parameters injected via YAML.
    \item A structured way in which simulation codes can be coupled together, with the outputs of one fed into another based upon the workflow logic.
    \item Usage of a standardised technology which is well documented and supported. Therefore use-case owners can enjoy a wealth of documentation and tutorials when developing their own CWL workflows.
\end{itemize}

\begin{figure}[h]
  \centering
  \includegraphics[scale=0.67]{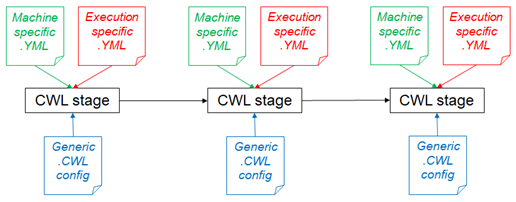}
  \caption{Illustration of CWL workflow on HPC machine coupling execution with configurations injected}
  \label{fig:cwl_workflow}
\end{figure}

The interplay between CWL and configurations is illustrated in Figure \ref{fig:cwl_workflow}, where a generic CWL configuration file is provided to the reference runner tool which defines the inputs and outputs for the file, along with the coupling of stages, but does not contain the concrete values. These are defined once for a simulation code and do not change between different scenarios or from machine to machine because, furthermore, a machine and execution specific YAML file is injected in. The idea is that not only can configuration options required from one execution of the code to the next be specialised, but furthermore configurations for different HPC machines exist which enable specialisation of the code between supercomputers in a portable manner. It can be thought of as the generic CWL configuration providing a skeleton which is then \emph{fleshed out} by the YAML files. Listing \ref{lst:workflow} provided an example of placing configuration data on the HPC machine before simulation execution, and commonly this configuration is in the form of YAML file(s) which concrete the generic CWL configuration.

\section{Case-study: Interactive urgent space weather ensembles}
\label{sec:casestudy}
The study of space weather \cite{sputnipic} involves modelling magnetic reconnection under several different condition. This involves a study of the configuration of the Earth’s magnetotail, can be applied for studying magnetic reconnection under different conditions, and is important because it is the magnetotail which protects the Earth and orbiting bodies from solar emissions. Phenomena in this magnetotail results in expensive satellite electrical failures and can also lead to electrical storms that short out earth-bound power networks.

The magnetic reconnection simulation code works on the basis of ensemble modelling, where many distinct permutations are executed and the results integrated. However HPC machines are often not suited for scheduling large numbers of individual jobs, for instance in our case when interactively simulating space weather, as the batch schedulers often tend to work in units of nodes. Consequently if the simulation is executing a low number of cores per ensemble then, on a machine with large numbers of cores per node such as ARCHER2 the UK national supercomputer, this can result in significant wasted resource. 

CWL provides two benefits which can ameliorate this problem; firstly a scatter mode which creates a number of concurrent workflow stages across a node and waits for these to complete. In the case of ensembles, these stages will be running the same executable, but with different parameters spread across the cores of a node. Secondly it provides a choice around granularity between the HPC machine and VESTEC marshalling and control system that was discussed in Section \ref{sec:marshallcontrol}. For example, if there are numerous simulations that must be coupled together then there is a choice of the marshalling and control system driving each of individually, submitting the next job when the previous one completes, and this is the finest grained approach. At the other extreme there is the coarse-grained view, where a single CWL workflow can be written which itself couples the jobs and is submitted to the HPC machine. As the execution of each CWL workflow is atomic, as far as the marshalling and control system is concerned, then the entire workflow will run through to completion before the marshalling and control system is notified it has finished. 

To explore the performance benefits of CWL scatter against batch queuing all ensembles separately, a synthetic benchmark has been developed. This provides a configurable number of ensemble members, each with a single-core job and we ran an experiment on ARCHER2 which imposes a maximum of 64 jobs queued at any one time and 16 jobs running concurrently. Due to this limitation, as the number of ensembles was increased, using the batch queue-only approach the marshalling and control system was forced to queue up submissions for the HPC machine and submit these only when an existing job had completed and left the queue. Otherwise the 64 limit would be reached and job rejected by ARCHER2. Each individual job in this benchmark is effectively a no-operation, completing immediately. By comparison, the CWL scatter mode will schedule ensemble members across the cores of nodes, enabling sharing of a node between many ensembles. Even though there is still a separate submission for each node, the overall number of nodes is greatly reduced as all the cores of a node are utilised running a member.

\begin{figure}[h]
  \centering
  \includegraphics[scale=0.67]{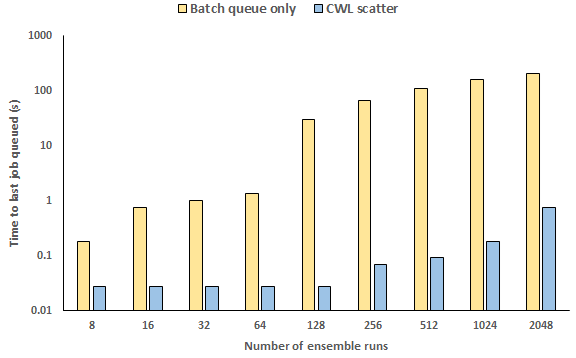}
  \caption{Time to last job being queued on HPC machine (ARCHER2) based on total number of ensembles and whether this is driven by the marshalling and control system (fine grained) or CWL workflow (coarse grained)}
  \label{fig:synthetic_ensemble}
\end{figure}

Figure \ref{fig:synthetic_ensemble} illustrates the time to the last job being queued on the HPC system for this synthetic benchmark, where the Y axis is log scale. This metric was adopted as it measures the time to interact with the HPC system rather than for jobs to run on the machine which depends upon a simulation code by simulation code basis. The major limitation of the batch queue only approach is that as each ensemble can only exploit one individual core, there are therefore very many queue submissions required, which equals the number of ensembles. Conversely, for the CWL scatter approach as CWL can scatter ensembles over the cores of a single node, there are 128 ensembles per node. Therefore, whilst there is still an individual submission required for each node, now the number of nodes needed is the number of ensembles divided by 128. It can be seen from Figure \ref{fig:synthetic_ensemble} that, as the number of ensembles is scaled, the completion time of the batch queue only approach is very significantly higher than that of the CWL scatter approach, most importantly because at 2048 ensembles batch queue only approach must submit 2048 separate jobs with the marshalling and control system having to queue up and track the completion of each and each progress through the queue, whereas the CWL scatter approach only needs to submit 16. It can be seen with the batch queue only approach how there is a sharp jump at 128 ensembles, this is because there is a maximum of 64 jobs in the batch queue and-so beyond that number of jobs the marshalling and control system must wait for existing jobs to complete before subsequent ensemble jobs can be submitted.

The experiment presented in Figure \ref{fig:synthetic_ensemble} is rather extreme, and the synthetic benchmark is a little artificial. Many, but not all, HPC systems provide job launching capabilities that enable the ability to run different executables across the cores of nodes. However, the downside is that often these can be fairly complex to interact with, driven by multiple issues of the launch command in a loop, and require support from the batch queue system. Such approaches can add complexity when preparing the job and limit the ability to target many different HPC machines irrespective of exactly what their batch queue system supports. From the marshalling and control system's perspective it is desirable for the workflow to be machine agnostic and not have to be concerned with such specific details. By contrast, the settings for the CWL scatter approach can be provided via machine specific YAML which is created when the use-case was installed on the machine, and/or by higher level parameters sent from the marshalling and control system. Nevertheless, this experiment illustrates one of the challenges faced when designing our general approach, namely the diversity of the urgent applications that we aim to support. This means that there are many different possible usage modes, ranging from large numbers of ensembles illustrated in Figure \ref{fig:synthetic_ensemble}, to single distributed memory applications that run over large numbers of cores. Consequently, one single approach to all these possibilities is not appropriate and instead selecting technologies that enable flexibility was required, and this has been demonstrated with the use-case implementations. Thus, giving the use-case developer a choice around key aspects by designing flexibility and generality into the approach is highly beneficial. 

\begin{figure}[h]
  \centering
  \includegraphics[scale=0.67]{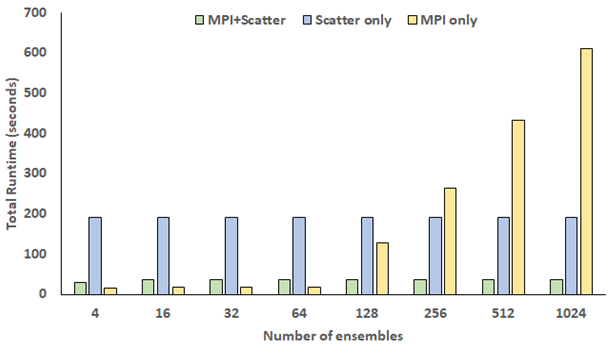}
  \caption{Total runtime of multiple B0z0.0 ensembles from space weather simulation on ARCHER2 as the number of ensembles is scaled and different approaches to scheduling these adopted.}
  \label{fig:space_weather}
\end{figure}

Considering interactive space weather prediction as a motivation, there is significant flexibility around scheduling the required ensembles and parallelism (i.e. number of cores) allocated to each individual member. We therefore undertook an experiment across space weather simulation ensembles on ARCHER2 where each member comprises 8 cores (which is optimal for the problem size being studied), with each core running an MPI process. Figure \ref{fig:space_weather} presents the results of this performance experiment where three configurations were tested; \textbf{MPI+Scatter} which uses a combination of MPI parallelism for each ensemble (8 MPI processes per ensemble) and CWL scatter to run 16 of these ensembles per node. \textbf{Scatter only} which is using CWL scatter in isolation, where each ensemble member is running over one core only and 128 ensembles per node. Lastly \textbf{MPI only} using the batch-queue system to schedule jobs only so there is one ensemble per node, but with MPI parallelism enabled for each ensemble member and hence each member is running over all 128 cores of a node.

It can be seen from Figure \ref{fig:space_weather} that the approach we have adopted is generally most efficient, especially for larger ensemble sizes. The Scatter only approach is uniformly consistent and always slower than the MPI+Scatter approach. This is because it only allocates one core per ensemble and-so is dominated by the runtime that this configuration results in. The MPI only approach is fastest for smaller numbers of ensembles because all 128 cores of the node are allocated to each ensemble member rather than the 8 cores as used by the MPI+Scatter approach. However, there is a trade-off, namely that with 8 cores per ensemble it is possible to run 16 ensembles per node compared with one ensemble member per node when allocating all 128 cores per ensemble. Consequently, for the MPI only case as the number of ensembles increases, especially after 64 ensembles where the marshalling and control system must queue these up and wait for others to complete due to limits imposed by the batch queue system, then the overhead of queuing up ensembles jobs and tracking them outweighs the benefits gained by running each member over 128 cores compared with 8 cores. This behaviour drives the very significant increase in overall runtime for the MPI only approach beyond 64 members, as the marshalling and control system must wait for ensemble jobs to complete before queuing up new ones. In consequence, for 1024 ensembles the MPI+Scatter approach requires the allocation of 64 nodes in total which can all fit as a single submission into the queue system, whereas the MPI only approach requires the allocation of 1024 nodes and this must be undertaken in segments. 

The purpose of the experiments described in this section have been to illustrate the flexibility provided by workflows for urgent, interactive, use-cases. Whilst it has been highlighted that some batch queue systems provide their own functionality to achieve similar scheduling flexibility, the benefit of exploiting this via CWL is that such decisions are undertaken in a standard manner across machines, irrespective of exactly what the batch queue system supports or not, with the marshalling and control system system only needing to provide numeric settings via YAML files in combination with any pre-defined machine specific configurations also in YAML. 

\section{Conclusions}
\label{sec:conclusions}
In this paper we have explored the role of workflows for expressing and running interactive urgent workloads on HPC machines. With two separate workflow systems, one in the VESTEC middleware system which provides marshalling and control functionality, and the other on the individual HPC machines, not only do these provide complimentary support for different facets but furthermore can work together to provide additional flexibility and most effectively suit the codes being run.

Using the space weather prediction urgent workload we then explored the challenges around ensemble scheduling on modern HPC machines, where we demonstrated that the flexibility provided by the two workflow systems working in combination is a powerful enabler for such situations. For future work we believe that it would be worthwhile to integrate the CWL and marshalling and control workflows, not necessarily at the underlying technology level but instead exploiting some of the safety provided by CWL definitions to ensure that those workflows expressed in the marshalling and control system are suitable. 

\subsubsection{Acknowledgements} 

The research leading to these results has received funding from the Horizon 2020 Programme under grant agreement No.\ 800904. This work used the ARCHER2 UK National Supercomputing Service (https://www.archer2.ac.uk). For the purpose of open access, the author has applied a Creative Commons Attribution (CC BY) licence to any Author Accepted Manuscript version arising from this submission.

%
%
%
%

\end{document}